\documentclass[aps,prl,twocolumn,superscriptaddress,amsmath,amssymb,showkeys]{revtex4-1}

\usepackage{graphicx}
\usepackage{xcolor}
\usepackage{fix-cm}
\usepackage{dcolumn}
\usepackage{bm}
\usepackage[normalem]{ulem}

\usepackage[utf8]{inputenc}
\usepackage[T1]{fontenc}
\usepackage{mathptmx}
\usepackage{etoolbox}
\usepackage[normalem]{ulem}
\begin{document}
\title{Pushing the Frontiers of Light: Magnetized Plasma Lenses and Chirp Tailoring for Extreme Intensities }

\author{Trishul Dhalia}
\email{trishuldhalia@gmail.com}
\author{Rohit Juneja}
\author{Amita Das}
\email{amita@iitd.ac.in}
\affiliation{Department of Physics, Indian Institute of Technology Delhi, Hauz Khas, New Delhi 110016, India
}

\begin{abstract}
In this work, an innovative scheme is proposed that exploits the response of magnetized plasmas to realize a refractive index exceeding unity for right circularly polarized (RCP) waves. Using two- and three-dimensional Particle-in-Cell (PIC) simulations with the OSIRIS 4.0 framework, it is shown that a shaped magnetized plasma lens (MPL) can act as a glass/solid state-based convex lens, amplifying laser intensity via transverse focusing. Moreover, by integrating three key ingredients, a tailored plasma lens geometry, a spatially structured strong magnetic field, and a suitably chirped laser pulse,  simultaneous focusing and compression of the pulse has been achieved. The simulations reveal up to a 100-fold increase in laser intensity, enabled by the combined action of the MPL and the chirped pulse profile. With recent advances in high-field magnet technology, shaped plasma targets, and controlled chirped laser systems, this approach offers a promising pathway toward experimentally reaching extreme intensities.
\end{abstract}
\maketitle

High power laser systems (in the exawatt range and beyond) are highly desirable for addressing several frontier problems in physics, such as electron - positron pair production \cite{di2012extremely,fedotov2023advances} and the study of quark-gluon plasmas \cite{mclerran1986physics,muller1985physics}. 
The advent of high-intensity lasers has created a pressing demand for optical systems capable of withstanding extreme power levels while still enabling precise control and manipulation of light. In this context, plasma emerges as an ideal optical medium, offering exceptional resilience where conventional solid-state optics would fail. Being already ionized, plasma can tolerate much higher intensities compared to the ($\sim10^{13}\text{W/cm}^{2}$)\cite{stuart1996optical,gamaly2002ablation} damage threshold of typical solid-state materials.
A variety of techniques have been developed to enhance laser intensity. Among them, Chirped Pulse Amplification (CPA) stands out as an ingenious method that overcomes the limitations of solid-state-based optics while still utilizing them, enabling the achievement of intensities in the range of ($\sim 10^{18} $ to $\sim 10^{23} \text{W/cm}^{2}$) \cite{strickland1985compression}. Spatiotemporal control of laser pulses has also been achieved with the help of plasma optics\cite{riconda2023plasma,li2024spatiotemporal}. The refractive index of the plasma can be tuned by the electron density $(n_e)$ for the incoming laser frequency $(\omega_l)$. Plasma lens configurations have been explored to achieve high intensities through transverse focusing of laser pulses \cite{ren2001compressing}. Additionally, laser pulse chirping has also been utilized for temporal compression using plasma mirrors, further contributing to intensity enhancement \cite{edwards2022plasma}.
Curved plasma geometries have recently gained attention for achieving extreme laser intensities. In a curved relativistic mirror (CRM), temporal compression and focusing of Doppler-upshifted light can reach the highest intensities with minimal spot size \cite{landecker1952possibility, bulanov2013relativistic, vincenti2019achieving, quere2021reflecting, kormin2018spectral}. While plasma mirrors require petawatt-class lasers and high contrast, a simpler and more robust approach would be to use long, high-energy (joule-level) chirped pulses \cite{tummler2009high, fisch2003generation, nagy2021high} and compressing them to relativistic intensities over ultrashort ranges, with easier control of contrast and spatiotemporal properties. This has been demonstrated using overdense plasmas with density gradients \cite{hur2023laser}. Other promising methods include Raman amplification of long, high-energy pulses \cite{trines2011simulations} and grating-based holographic plasma lenses \cite{edwards2022holographic, leblanc2017plasma}, underscoring the potential of plasma optics to surpass current laser intensity limits.

Building on these advances, a particularly promising frontier involves combining intense laser–plasma interactions with strong magnetic fields for novel optical applications. Magnetized plasmas have already been proposed for a variety of phenomena, including resonant energy absorption \cite{PhysRevE.110.065213,Juneja_2023,vashistha2020new,juneja2024enhanced}, magnetic transparency \cite{mandal2021electromagnetic,goswami2021ponderomotive}, and high harmonic generation \cite{dhalia2023harmonic,maity2021harmonic}. A recent study on magnetized low-frequency (MLF) scattering has been used to amplify laser pulses in strong magnetic fields \cite{edwards2019laser}.


Generating magnetic fields of the required strength for these applications, on the order of tens of kilo-Teslas (kT), remains a major experimental challenge and is beyond the capabilities of currently available technologies. However, recent breakthroughs have demonstrated fields at the kT level \cite{nakamura2018record, choudhary202510}, and emerging proposals suggest that Megatesla (MT) fields may soon be achievable. In particular, target geometry-driven mechanisms have even indicated the possibility of generating Giga Gauss level fields ($\sim 0.1 MT$) \cite{pan2025gigagauss, korneev2015gigagauss}. These advances point toward a promising convergence of plasma-optics techniques and high-field magnetized environments, which could unlock unprecedented opportunities for controlling, compressing, and amplifying laser pulses, thereby pushing the frontiers of high-power laser science.

\begin{figure}
  \centering
  \includegraphics[width=8cm]{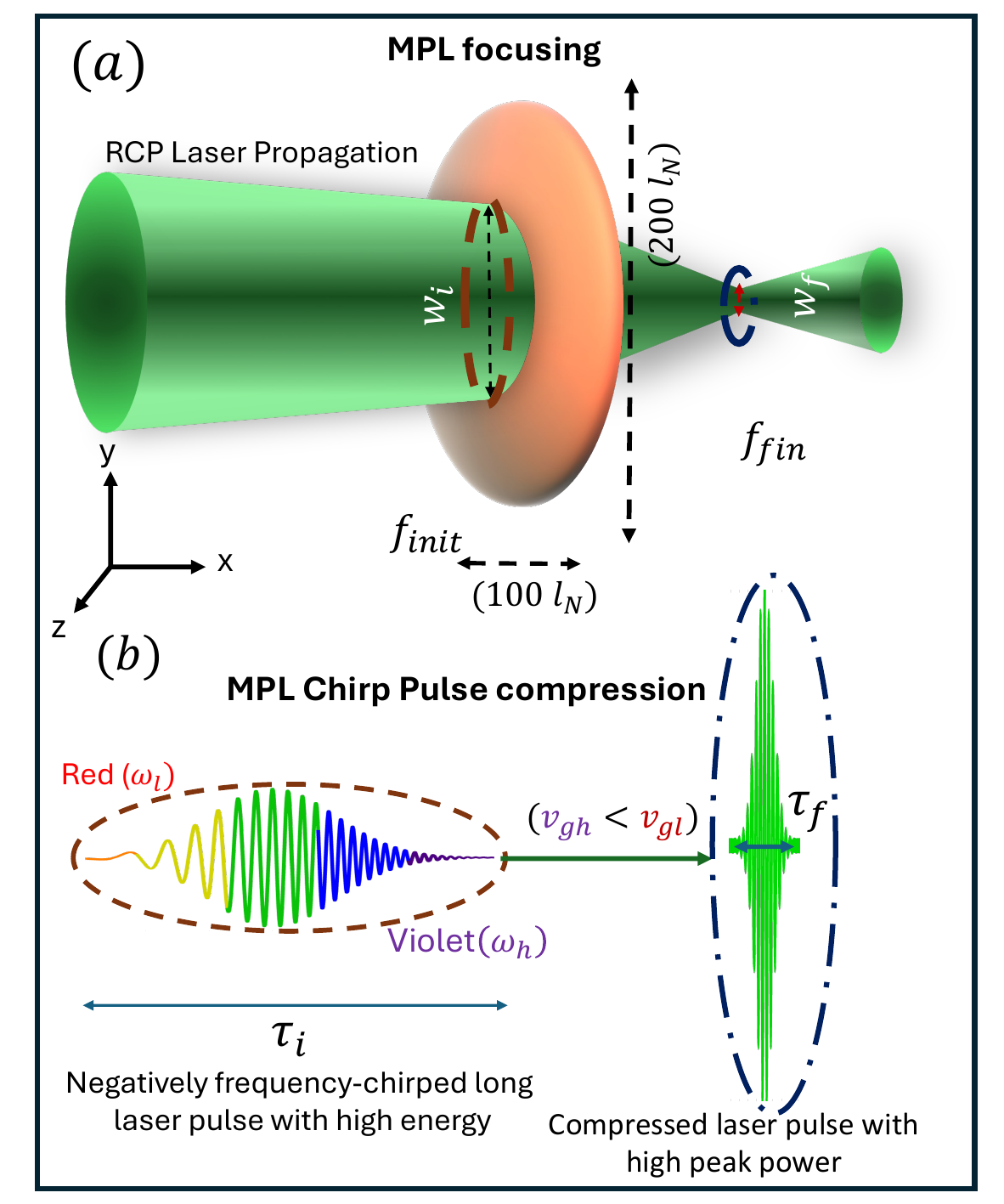}
  \caption{The figure demonstrates the schematic representation (not to scale) of the geometry chosen for 3D simulation. Subplot (a) shows the negatively chirped right circularly polarized laser pulse incident onto a magnetized plasma lens (MPL) immersed in an inhomogeneous magnetic field $\vec{B}_{ext}(x,y,z)$. Here $w_i,w_f$ represent incident and final spot sizes, and $f_{\text{init}},f_{\text{fin}}$ denote the location of the incident and final focus points of the field. Subplot (b) demonstrates simultaneous chirp pulse compression due to inhomogeneous $\vec{B}_{ext}$ in MPL. Here $\tau_i, \tau_f$ represent the pulse duration of the incident chirped pulse and the final compressed pulse after passing through the MPL.}
\label{fig:schematic}
\end{figure}

In this Letter, we propose a novel approach to reach extreme intensities based on chirp pulse compression as well as self-focusing of a long, negatively chirped, wide spot right circularly polarized laser through an underdense magnetized convex-shaped plasma lens (MPL). With the 3D and 2D particle-in-cell (PIC) simulations, we show that reaching extreme intensities is possible using MPL. The schematic of this scheme has been shown in Figure \ref{fig:schematic}(a). 

While curved mirror geometries have been extensively studied in plasma optics \cite{vincenti2019achieving,quere2021reflecting}, curved lens geometries have attracted less attention. This is mainly because, even in an underdense plasma where the laser frequency $\omega_{l}$ exceeds the plasma frequency $\omega_{pe}$, allowing optical transmission, the refractive index of an unmagnetized plasma remains less than unity. Consequently, the phase velocity exceeds the speed of light $(v_p > c)$, preventing the convergence of incident rays and causing laser pulses to diverge from a convex plasma lens \cite{dhalia2025laser}.

However, the introduction of an external magnetic field adds a crucial tuning parameter: the electron cyclotron frequency $\omega_{ce}$. Along with electron density and laser frequency, this enables precise control over the plasma’s refractive index. Under sufficiently strong magnetic fields, the magnetized response of plasma species can increase the refractive index beyond unity, allowing the plasma to mimic conventional optical materials like glass or solids. This opens the possibility of using magnetized plasma lenses for focusing high-power laser pulses without the typical power threshold limitations.



When the magnetic field is aligned  along the laser propagation direction,  the dispersion relation for the R-mode is given by  \cite{kruer1988physics,chen1984introduction},
 \begin{equation}
     n_R^2 =\frac{c^2}{v_p^2} =1-\frac{\omega_{pe}^2/\omega_{l}^2}{(1-\omega_{ce}/\omega_{l})}
\end{equation}
For the case  $(\omega_{ce}>\omega_{l})$, the refractive index  $(n_R >1)$, results in the phase velocity that is lower than the speed of light within the plasma. Consequently, a magnetized plasma can act as a convex lens, exhibiting focusing properties similar to those of a conventional converging lens. Interestingly, this simple yet profound characteristics of magnetized plasma as a convex lens has remained unexplored until recently \cite{dhalia2025laser}. 

We explore this property to achieve a higher laser field intensity. When the wavelength $(\lambda)$ of the incident light pulse is much smaller than the aperture size $(\lambda <<D)$, the long laser pulse can be approximated as a ray. Under this approximation the  focal length $f$ of such a convex magnetized plasma lens can be estimated as   
 \begin{equation}\label{lensmaker}
     \frac{1}{f_{\text{fin}}}=(\bar{n}_R-1) \{ \frac{2}{R}+\frac{(\bar{n}_R-1)L}{\bar{n}_RR^2} \}
 \end{equation}
Here, $R$ is the radius of curvature of the convex plasma lens, and $L$ is the thickness of the plasma. Here, $\bar{n}_R$ is the average refractive index of the plasma lens. This formula is valid as long as the plasma does not absorb any laser energy and behaves like an ideal, thick, solid lens. For the condition  $(\omega_{ce}>\omega_{l})$,  the laser does not encounter any resonance inside the plasma, resulting in negligible energy absorption for collisionless cases. However, if significant energy is absorbed by the plasma, the focal length predicted by the  lens maker's formula may deviate from the actual value. 

While the magnetized lens is used for transverse compression of the laser pulse to enhance its intensity, simultaneous temporal compression is achieved by appropriately choosing the chirp of the original laser beam 
\cite{jha2014evolution}.
The group and phase velocity of an RCP wave under R-mode geometry is given by \cite{dhalia2023harmonic}, 
\begin{equation}\label{phase}
    v_p(\omega,\omega_{ce})=\frac{\omega}{k}=c\sqrt{\frac{\omega(\omega_{ce}-\omega)}{\omega_{pe}^2+\omega_{ce}\omega-\omega^2}},
\end{equation}
\begin{equation}\label{group}
    v_g(\omega,\omega_{ce})=\frac{d\omega}{dk}=\frac{2c^2k(\omega_{ce}-\omega)}{c^2k^2+\omega_{pe}^2-3\omega^2+2\omega_{ce}\omega}
\end{equation}    
These expressions show that both group and phase velocities decrease as the frequency  increases. As a result, launching a negatively chirped long laser pulse, where the high frequency  $(\omega_{h})$ leads  in time and the low-frequency component $(\omega_{l})$ trails, through the MPL enables pulse compression. The difference in group velocities cause the frequency component to converge, leading to a compressed pulse. For a chirped laser pulse, spanning the frequency range  $(\omega_{l},\omega_{h})$, the required pulse duration $(\Delta \tau)$ for maximal compression is given by, 
\begin{equation}
    \Delta \tau =\frac{L}{v_{gh}}-\frac{L}{v_{gl}}.
\end{equation}
Here, $v_{gh}$ and $v_{gl}$ are the  group velocities corresponding to $\omega_h$ and $\omega_l$ respectively, and $L$ is the plasma length along the laser axis. The duration $\Delta \tau$  is chosen such  that these two frequency components reach the desirable location, typically the plasma edge, for optimal compression, simultaneously.  However,  defining the chirp profile  $\omega (\tau)$ within the interval $(\Delta \tau)$ is equally important. For a linear chirp,  $\omega(\tau) =\omega_0 -\alpha \tau $, only the extreme frequencies overlap at the plasma edge, while the intermediate frequencies arrive at different times due to the nonlinear dependence of  the group velocity  on frequency Eq.\ref{group}. This temporal mismatch leads to an extended focal region.  To achieve optimal compression, a nonlinear chirp is therefore essential. Such profiles can now be realised through ultrafast pulse shaping techniques \cite{weiner2011ultrafast,tull1997high,fetterman1998ultrafast,goswami2003optical,wang2011laser}.  The required  chirp profile $\omega( \tau) $ is obtained by inverting 
  \begin{equation}\label{chirp_prof}
      \tau(\omega) =\int_{x_1}^{x_2}\frac{dx}{v_g(\omega,\omega_{ce})}, 
  \end{equation}
  where $ v_g (\omega, \omega_{ce}) $ is the group velocity in  the magnetized plasma and  
 $L=x_2-x_1$ is the lens length. In our case, the cyclotron frequency varies linearly as, $\omega_{ce}(x)=\omega_{ce_0}-p\cdot x$ with $\omega_{ce,0}=eB_0/m_e$ and $p$  characterizing the field gradient.

With this straightforward methodology in mind, we performed a proof-of-principle three-dimensional (3D) PIC simulation with the massively parallel OSIRIS-4.0 PIC code \cite{fonseca2002osiris,davidson2015implementation}. The code uses conventional normalized variables and fields, wherein the length is normalized by the electron skin depth $c/\omega_{pe}$ and time by the electron plasma period $\omega_{pe}^{-1}$ [see Appendix: A for more details]. 

 
 An external magnetic field linearly decreasing along the x-axis, having the following normalized form $B_{N,ext} (B_0, \delta)$ has been applied: 
\begin{multline}
  \label{eq:mag_field}
    \vec{B}_{N, ext}(B_0,\delta) = \frac{\vec{B}_{ext}} { (m_ec \omega_{pe}/e)}  = [B_0 -\delta(x-110)]\hat{i} \\
    + [0.5\cdot\delta(y-110)]\hat{j} +  [0.5\cdot\delta(z-110)]\hat{k}   
\end{multline}
The choice of the expression in Eq.(\ref{eq:mag_field}) ensures that the $\nabla \cdot \vec{B}_{ext} = 0$. We have fixed the value of the parameter $\delta =0.005$ and have varied the strength of the magnetic field through $B_0$ in our simulations. For  $B_0 = 2.2$, the value of  $\vec{B}_{ext}$ is chosen to range between $\sim 15$  to $25$ kT for an $800 nm$ laser and $\sim 2$  to $4$ kT for $CO_2$ laser [$\lambda= 9.42\mu m$].  

For definiteness considering the  $(\lambda=800 nm)$ laser, having a total energy of $0.63 mJ$, pulse duration $\approx 80 fs$ and spot-size $\approx12 \mu m$, corresponds  to an intensity of  $I \approx 1.36\times 10^{16} \text{W/cm}^{2}$ and the normalized vector potential amplitude $a_0 \approx$  0.08 = 0.85 $\sqrt{I(10^{18}\text{W/cm}^{2})}\lambda (\mu m)$. The laser energy in the simulation has been chosen to be small ($\sim mJ$) to comply with the constraint of available computational resources. It is feasible to scale it to several joules by extending the pulse duration to \textit{pico-} or \textit{nano}seconds regime, and enlarging the lens dimensions to $\sim cm$ scale. The focusing element is a convex plasma lens, consisting of fully ionized hydrogen plasma [see figure \ref{fig:schematic}], with electron density  $n_e=0.69 n_c$. Here $n_c$ is the critical density for the central laser wavelength. Such a low density plasma can be produced experimentally using a foam target based on polymer aerogels \cite{rosmej2025advanced,khalenkov2006experience}. The pulse  is negatively chirped, with frequency nonlinearly varying within a range  $\omega_l \in (0.7,1.7)\omega_{pe}$. The nonlinear chirp employed in the simulation is derived from equation (\ref{chirp_prof}) for the external magnetic field $\vec{B}_{ext}(2.2,0.005)$ can be represented  by a  fourth-order polynomial:
\begin{equation}
\label{eq:nchirp}
    \omega_{non}(\tau) =\omega_0+\omega_1\tau+\omega_2\tau^2+\omega_3t^3+\omega_4\tau^4, 
\end{equation}
with coefficients  $\omega_0=1.5276,\omega_1=0.0028,\omega_2=4.5177\times 10^{-6}, \omega_3=5.1483\times 10^{-7}, \omega_4=-1.0223\times 10^{-8}$. This profile is employed in simulations with a nonlinear chirp. 

\begin{figure}[!ht]
    \centering    \includegraphics[width=8cm]{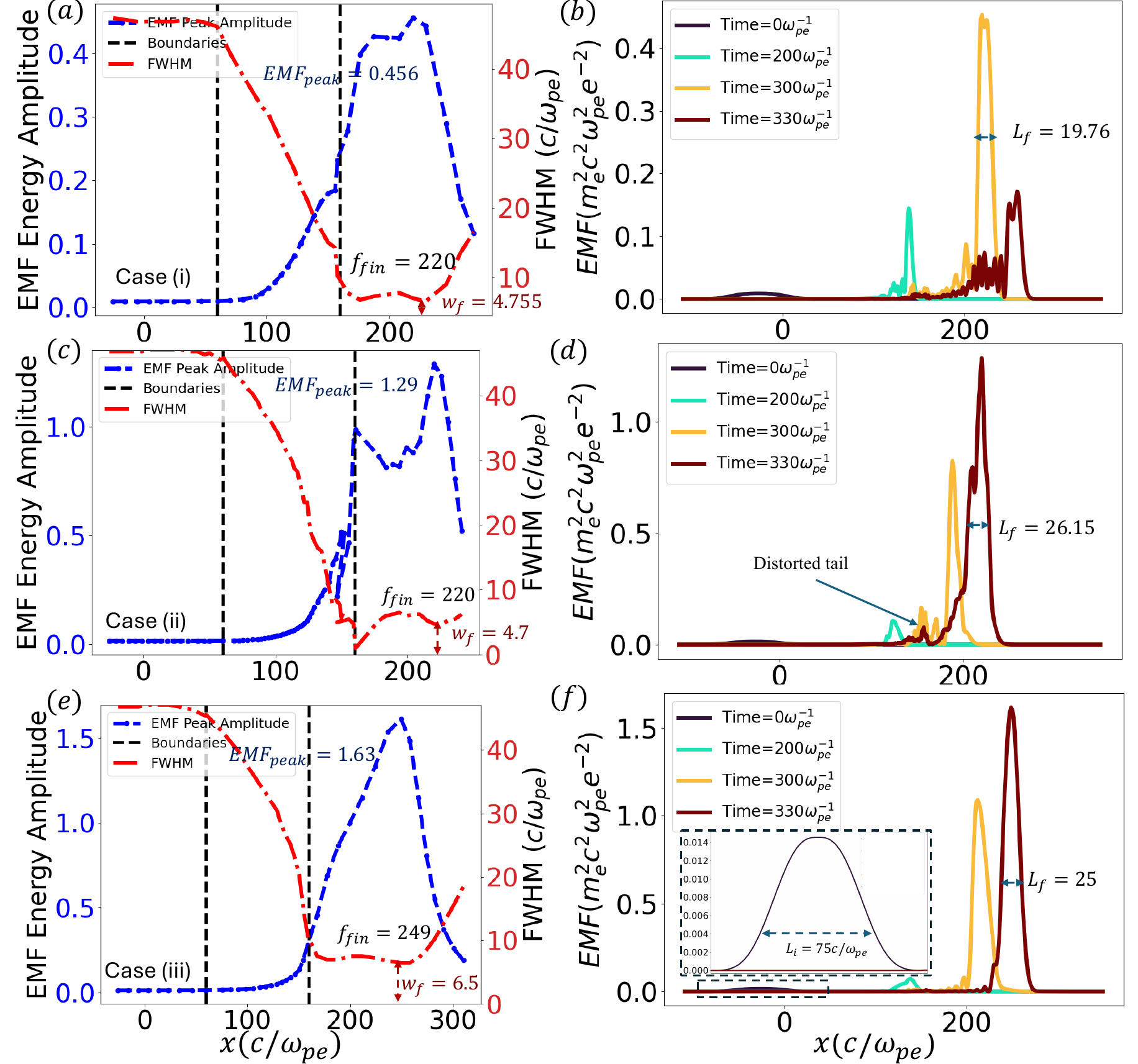}
    \caption{The peak of EMF energy density with simultaneous transverse spot-size (FWHM) evolution for a linear chirp laser has been shown in figure a) for a convex lens immersed in the magnetic field for $\vec{B}_{N,ext}(2.2,0.005)$ described in equation (\ref{eq:mag_field}). Figure (b) depicts the longitudinal profile of the laser at different times, and peak amplitude is achieved at $t=330\omega_{pe}^{-1}$. Figures (c,d) are similarly plotted for the same plasma lens and $\vec{B}_{N,ext}(2.2,0.005)$ but with the nonlinear chirp laser. Figure (e,f) is plotted for the same plasma lens and a little higher $\vec{B}_{N,ext}(2.4,0.005)$ but with a linear chirp profile.  }
    \label{fig:fwhm_emf_energy_liner_non_linear_const_B}
\end{figure}

\begin{figure*}[!ht]
    \centering    \includegraphics[width=16cm]{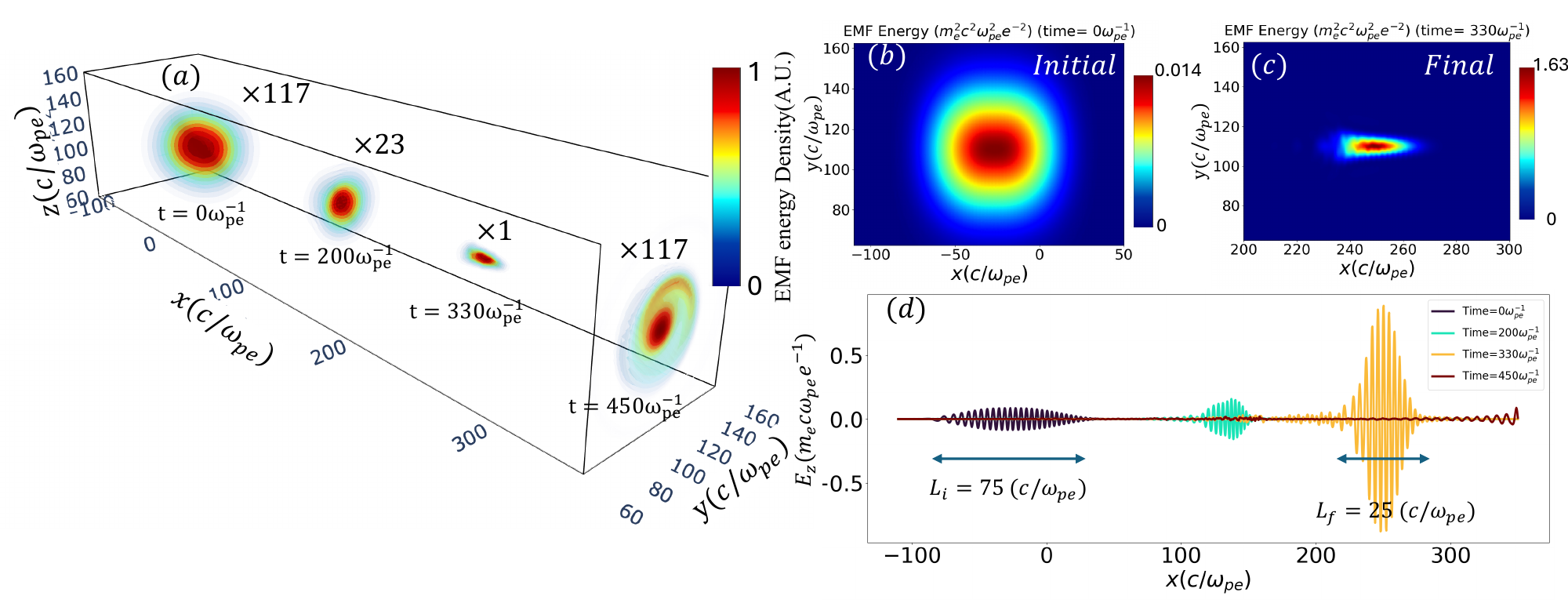}
    \caption{The figure $(a)$ demonstrates the time evolution of an incoming laser pulse launched with a peak amplitude of $0.014 m_e^2c^2\omega_{pe}^2e^{-2}$ ($\times 117$ zoom) at $t=0 \omega_{pe}^{-1}$  and a maximum compressed and focused pulse achieved at a location  $f_{\text{fin}}=249c/\omega_{pe}$ at time $t=330 \omega_{pe}^{-1}$ with a peak amplitude reaching a $1.63 m_e^2c^2\omega_{pe}^2e^{-2}$ ($\sim 117 $ fold increase); afterwards it again diverges. The figure $(b,c)$ shows surface projection in the $x-y$ plane at the center $z=110c/\omega_{pe}$ of EMF energy density for the initial and final max compressed and focused pulse. Figure $(d)$ presents a  1-D snapshot of the longitudinal pulse profile at various times. This clearly shows that the chirp pulse is compressed by $1/3$ times from MPL.  }
    \label{fig:3D_EMF}
\end{figure*}

We present the observations of 3D simulation for three representative cases corresponding to (i) $B_ 0 = 2.2$ and a linear chirp profile. The cases   (ii) and (iii) are for $B_0 = 2.2$ and $B_0 = 2.4$, respectively with  a nonlinear chirp profile $ \omega_{non}(\tau)$ defined by Eq.(\ref{eq:nchirp}). It should be noted that this profile has been optimized for   $B_0 = 2.2$. 
According to the lens maker’s formula [Eq. (\ref{lensmaker})], the predicted focal distances, corresponding to the points of maximum intensity enhancement, are $f_{\text{fin}} \sim 216$ for cases (i) and (ii), owing to their identical $B_0$ values, and $f_{\text{fin}} \sim 246$ for case (iii) . The simulations yield closely matching results, as seen from the intensity peaks in the left-column subplots of Fig. \ref{fig:fwhm_emf_energy_liner_non_linear_const_B}. In these plots, the maximum EMF energy (blue dashed curve) and the full width at half maximum (FWHM) of the transverse spot (red dashed–dotted curve) are shown as functions of the axial distance $x$ for all three cases. The decrease in FWHM with increasing intensity confirms that transverse focusing is central to the observed enhancement. The vertical black dashed lines mark the axial extent of the plasma lens.
The right-column plots further illustrate the axial position of the EMF pulse at different times. The results show that maximum amplification indeed occurs at the focal location predicted by Eq. (\ref{lensmaker}). Moreover, the temporal pulse duration decreases as the intensity rises, indicating the role of chirp-induced pulse compression. It should be noted that in both cases (i) with a linear chirp profile, and case(ii) with the optimized chirp profile, a noisy focused pulse gets formed. Case (i), in fact, shows a broadened axial region where the focusing occurs.  In both these cases, one observes that the focusing being tighter near the lens, the pulse acquires significantly higher amplitude while it is still inside the plasma [see Appendix: B]. We note that the intensity in fact exceeds $10^{18}, \text{W/cm}^2$, which is known to trigger a nonlinear plasma response \cite{li2024spatiotemporal}, thereby making the pulse significantly noisy. A natural question is whether the outcome can be improved by shifting the focal point slightly outside the plasma lens. To test this, we adjust the magnetic field to $B_0 = 2.4$ in case (iii) of our simulations, while keeping all other parameters unchanged. This modification shifts the focal spot further away from the plasma lens. Consequently, there is virtually no amplification within the lens region. In this configuration, the pulse remains clean while exhibiting stronger amplification, making it the most favorable scenario. 

We now examine this best-case scenario, namely case (iii), in greater detail.
Figure \ref{fig:3D_EMF}(a) presents the time evolution of the EMF energy density for this case.  The chirped Gaussian laser pulse exhibits a significant amplification of EMF energy density ($2a_0^2$ in normalized units), increasing from an initial peak of $0.014$ [$I_{\text{init}} \approx 1.36 \times 10^{16}\text{W/cm}^2$, Fig. \ref{fig:3D_EMF}(b)] to a final peak of $1.63$ [$I_{\text{fin}} \approx 1.58 \times 10^{18}\text{W/cm}^2$, Fig. \ref{fig:3D_EMF}(c)] at $t = 330\omega_{pe}^{-1}$. This corresponds to nearly a two-order-of-magnitude gain in peak intensity. As the pulse propagates through the magnetized plasma lens, it undergoes both self-compression and self-focusing, reaching an optimally compressed and focused state at $f_{\text{fin}} = 249c/\omega_{pe}$. Beyond this focal point, the pulse gradually defocuses and exits the simulation domain. Importantly, no significant reflection losses are observed at the plasma boundaries, indicating efficient transmission through the lens. At $t = 330\omega_{pe}^{-1}$, Fig. \ref{fig:3D_EMF}(c) shows that the focused pulse remains clean and largely distortion-free. This stability results from the focal spot lying outside the plasma region and the large aperture of the lens ($D = 200l_N \approx 27\lambda \gg \lambda$), which suppresses diffraction from the curved plasma surface.

\begin{figure}[!ht]
    \centering  \includegraphics[width=8cm]{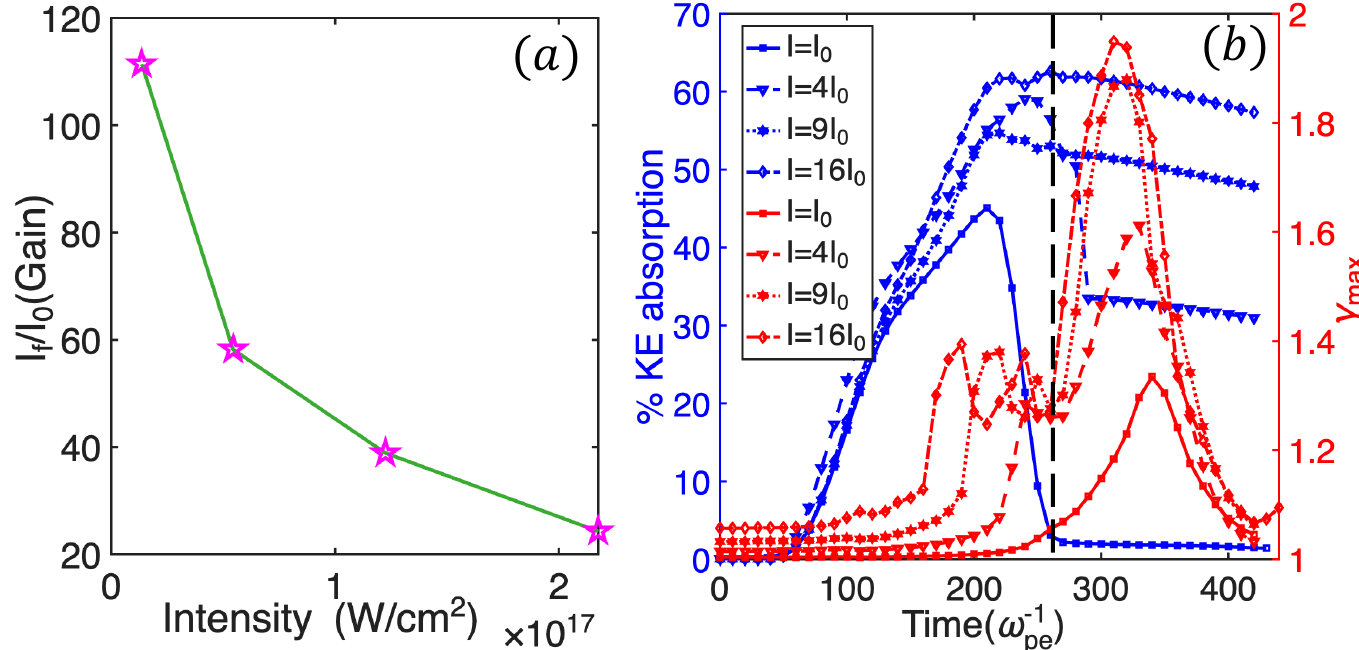}
    \caption{Figure (a) demonstrates the ratio of final and initial intensity as a function of initial intensity. (b) The kinetic energy absorbed by the electrons in plasma with time has been plotted here. The black dashed line represents the laser exiting time from the plasma lens.}
    \label{fig:intensity+absorption}
\end{figure}

To confirm that the increased intensity inside the plasma lens is responsible for the detrimental effect, we performed simulations with higher incident intensities while keeping all other parameters fixed. The results, shown in Fig. \ref{fig:intensity+absorption}, clearly demonstrate that amplification decreases with increasing incident energy, and in each case, intensity enhancement occurs within the plasma region. This behavior can be understood as follows: with higher incident intensity, the relativistic factor $\gamma_{\text{max}}$ increases, which in turn reduces the electron cyclotron frequency according to $\omega_{ce} = \omega_{ce0}/\gamma_{\text{max}}$. Thus, higher intensity drives $\gamma_{\text{max}}$ to values that satisfy the resonance condition $\omega_{ce} \approx \omega_{l}$ inside the plasma. Under such circumstances, irreversible resonant heating dominates, leading to a reduced net intensity gain $(I_f/I_0)$. A possible remedy is to choose a larger value of $B_0$, thereby shifting the focal point further outside the plasma lens. In fact, the parameter $B_0$ here defining the magnetic field profile turns out to be an important tuning parameter as shown in Appendix C. 

In conclusion, we propose a curved plasma-lens scheme that employs magnetized plasma optics to achieve large intensity gains from low-intensity, high-energy, and long-duration pulses. A negatively chirped Gaussian beam undergoes self-compression and self-focusing due to the combined effect of an inhomogeneous magnetic field and a curved plasma geometry, reaching its shortest duration and smallest waist at the plasma lens focus. Unlike curved mirror concepts in unmagnetized plasmas that demand petawatt intensities, our approach requires only modest initial intensities with large focal spots.

 \textit{Acknowledgements}:
 \indent The authors would like to acknowledge the OSIRIS Consortium, consisting of UCLA and IST (Lisbon, Portugal), for providing access to the OSIRIS 4.0 framework, which is the work supported by the NSF ACI-1339893. A.D. acknowledges support from the Anusandhan National Research Foundation (ANRF) of the Government of India through core grant CRG/2022/002782 as well as her J C Bose Fellowship grant JCB/2017/000055. The authors would also like to thank IIT Delhi HPC facility for computational resources. T.D. also wishes to thank the Council for Scientific and Industrial Research (Grant No. 09/086/(1489)/2021-EMR-I) for funding the research.

 \indent Authors report no conflict of interest

\bibliographystyle{apsrev4-1}

\bibliography{jpp-instructions.bib}

\bigskip
\newpage
\section{Appendix A: Simulation Details}
The 3-D simulations were run on a local machine with 96 cores for approximately  $\sim120$ hours ($\approx 11,520 $ core-hours),  using optimized convex lens geometry and a chirped right-hand Circularly polarised (RCP)  Gaussian laser in the presence of an external magnetic field.   Particle weighting used cubic interpolation on a staggered Yee grid, and the regular Yee solver was used to update the fields. 
 The simulation domain is discretized into $1840 \times 880 \times 880$ cells with spatial resolution of $dx=dy=dz= 0.25 c/\omega_{pe}$ and temporal resolution of  $dt = 0.05 \omega_{pe}^{-1}$.  
 
 The  plasma lens has a homogeneous density and is finite only within the region 
\begin{multline}
    n_e =n_0 \    \ , \Re  \in  c_0(\sqrt{y^2+z^2}-110)^2 +60  \\
    \leq x\leq -c_0(\sqrt{y^2+z^2}-110)^2 +160
\end{multline}
 Here, parameter $c_0$ determines the shape of the plasma and has been chosen as $0.005$. The radius of curvature for such a lens turns out to be $R=100 c/\omega_{pe}$. The incident laser pulse duration is  $ \Delta \tau =154  \omega_{pe}^{-1}$ and the spot-size has a diameter $80 c/\omega_{pe}$.  Thus the  spot size is considerably small compared to the transverse dimension of the plasma lens $D=200c/\omega_{pe}$. This minimizes the effect of spherical aberration \cite{li2022influence}. The total run time of the simulation was $460\omega_{pe}^{-1}$. 
 The dynamics of both electrons and ions have been considered in the simulation. 

\section{Appendix B: Evolution of EMF energy density for case (i) and case (ii)}
The Fig. \ref{fig:3D_figure_case_2_3} shows the time evolution of EMF energy density in cases (i) and (ii). Though, compression and focusing take place in both these cases. However, a comparison with Fig. \ref{fig:3D_EMF} for case (iii) in the manuscript shows that the quality of compression in these cases is not good. While case (i) shown in Fig. \ref{fig:3D_figure_case_2_3} shows elongated focus in the longitudinal direction,  for case (ii) the energy gets reflected from the edge of the lens and propagates inwards, thereby causing a loss of energy. 
\renewcommand{\thefigure}{S1}
\begin{figure*}[!ht]
    \centering
    \includegraphics[width=0.9\linewidth]{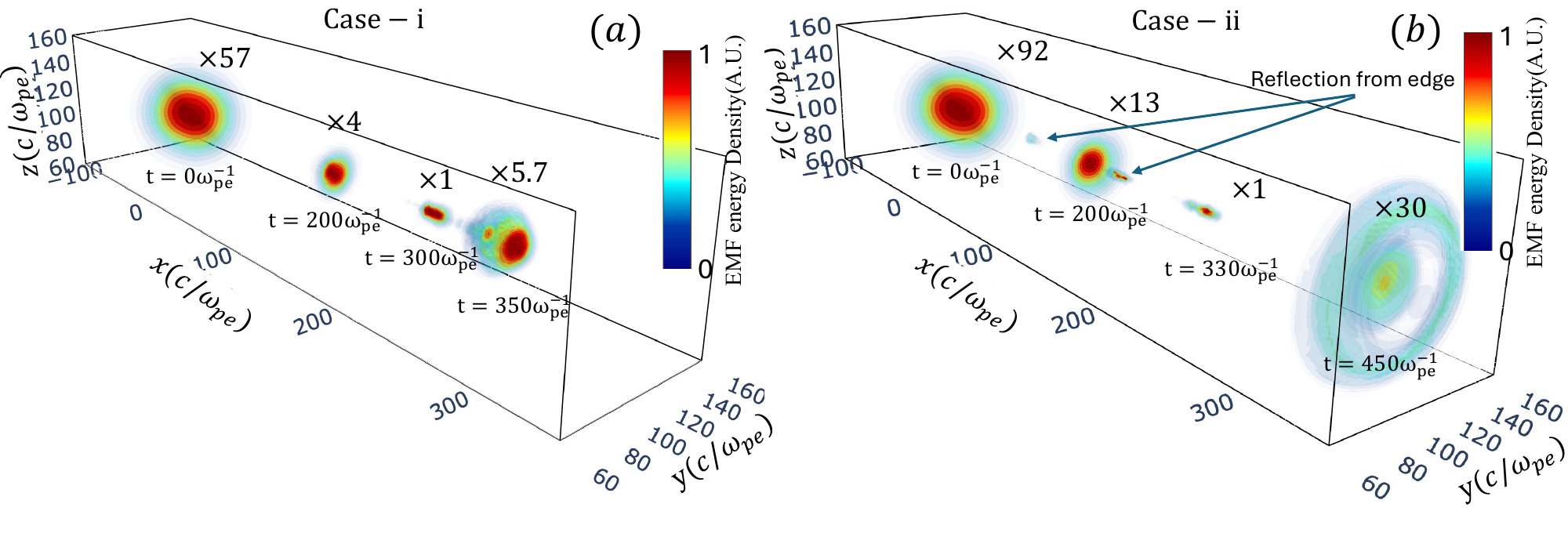}
    \caption{ Normalized EMF energy density evolution in time for cases I and II under interaction with MPL, shown in figures $(a)$ and $(b)$ respectively. At each time t, $(\times M)$ magnified field has been plotted.   }
    \label{fig:3D_figure_case_2_3}
\end{figure*}

\section{Appendix C: Amplification  by  tuning the parameter $B_0$}
Two-dimensional simulations for various values of the parameter $B_0$ have been carried out spanning the range  $B_0\in [2.2,2.6]$. It can be observed from Fig. \ref{fig:Peak_energy_mag}  that the best result is obtained for $B_0 = 2.4$. Thus, the profile of the magnetic field acts as an important tuning parameter.  

\renewcommand{\thefigure}{S2}
\begin{figure}[!ht]
    \centering  \includegraphics[width=8cm]{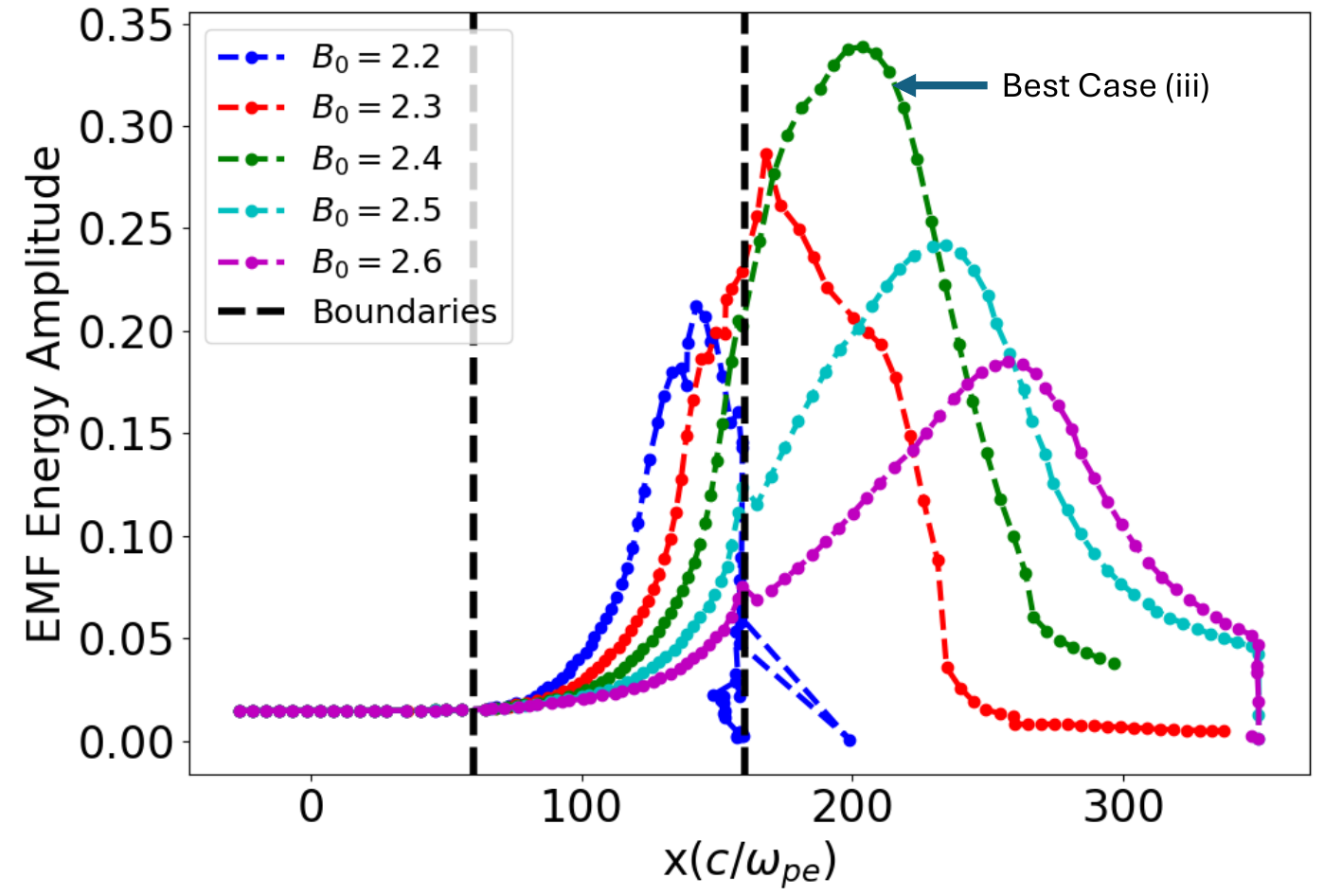}
    \caption{The figure shows the evolution of peak EMF energy amplitude for different cases of applied values of $B_0$ in $\vec{B}_{N,ext}$ performed with 2D PIC simulations.  }
    \label{fig:Peak_energy_mag}
\end{figure}

\end{document}